\documentclass[sigconf]{acmart}
\AtBeginDocument{%
  \providecommand\BibTeX{{%
    \normalfont B\kern-0.5em{\scshape i\kern-0.25em b}\kern-0.8em\TeX}}}

\copyrightyear{2023}
\acmYear{2023}
\setcopyright{rightsretained}
\acmConference[SA Conference Papers '23]{SIGGRAPH Asia 2023 Conference Papers}{December 12--15, 2023}{Sydney, NSW, Australia}
\acmBooktitle{SIGGRAPH Asia 2023 Conference Papers (SA Conference Papers '23), December 12--15, 2023, Sydney, NSW, Australia}
\acmDOI{10.1145/3610548.3618252}
\acmISBN{979-8-4007-0315-7/23/12}

\acmSubmissionID{998}

\usepackage{subcaption}
\usepackage{multirow}

\usepackage[title, toc, page]{appendix}

\usepackage[ruled]{algorithm2e} 

\SetAlFnt{\small}
\SetAlCapFnt{\small}
\SetAlCapNameFnt{\small}
\SetAlCapHSkip{0pt}

\usepackage{enumitem}
\usepackage{color}
\definecolor{blue}{rgb}{0,0,0.8}
\definecolor{green}{rgb}{0,0.6,0}
\definecolor{red}{rgb}{0.7,0,0}
\def\blue{\color{blue}}

\def\mat#1{\mathbf{#1}}

\newcommand{\mX}{\mat{X}}

\newcommand \mY	{\mat{Y}}
\newcommand \mZ	{\mat{Z}}

\newcommand{\real} {\mathbb{R}}

\newcommand{\ignore}[1]{}

\begin{document}

\title{MOCHA: Real-Time Motion Characterization via Context Matching}


\author{Deok-Kyeong Jang}
\email{shadofex@kaist.ac.kr}
\orcid{0000-0002-7567-4339}
\affiliation{%
  \institution{KAIST and MOVIN Inc.}
  \country{South Korea}
}

\author{Yuting Ye}
\email{yuting.ye@meta.com}
\affiliation{%
  \institution{Reality Labs, Meta}
  \country{USA}
}

\author{Jungdam Won}
\email{jungdam@imo.snu.ac.kr}
\orcid{0000-0001-5510-6425}
\affiliation{%
  \institution{Seoul National University}
  \country{South Korea}
}

\author{Sung-Hee Lee}
\authornote{Corresponding author}
\email{sunghee.lee@kaist.ac.kr}
\orcid{0000-0001-6604-4709}
\affiliation{%
  \institution{KAIST}
  \country{South Korea}
}


\begin{abstract}
Transforming neutral, characterless input motions to embody the distinct style of a notable character in real time is highly compelling for character animation. This paper introduces MOCHA, a novel online motion characterization framework that transfers both motion styles and body proportions from a target character to an input source motion. MOCHA begins by encoding the input motion into a motion feature that structures the body part topology and captures motion dependencies for effective characterization. Central to our framework is the Neural Context Matcher, which generates a motion feature for the target character with the most similar context to the input motion feature. The conditioned autoregressive model of the Neural Context Matcher can produce temporally coherent character features in each time frame. To generate the final characterized pose, our Characterizer network incorporates the characteristic aspects of the target motion feature into the input motion feature while preserving its context. This is achieved through a transformer model that introduces the adaptive instance normalization and context mapping-based cross-attention, effectively injecting the character feature into the source feature. We validate the performance of our framework through comparisons with prior work and an ablation study. Our framework can easily accommodate various applications, including characterization with only sparse input and real-time characterization. Additionally, we contribute a high-quality motion dataset comprising six different characters performing a range of motions, which can serve as a valuable resource for future research.  
\end{abstract}




\begin{teaserfigure}
  \centering
  \includegraphics[width=7.0in]{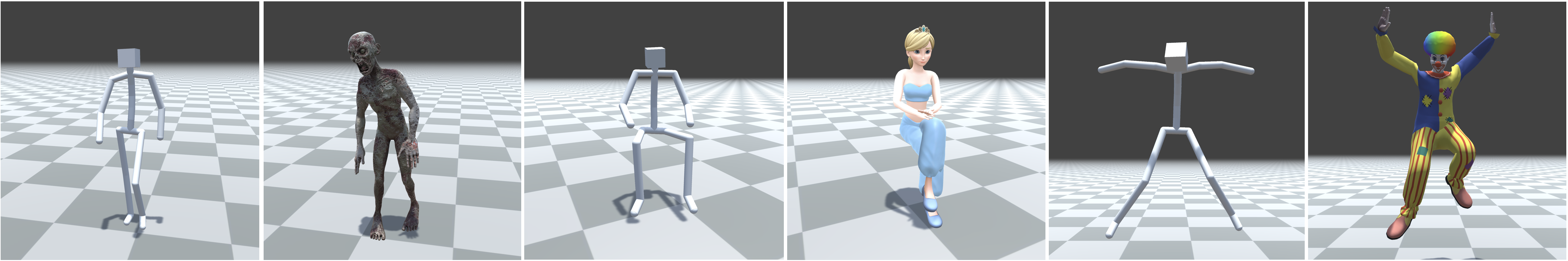}
  \caption{Our characterization framework transforms neutral motions to express distinct style of characters in real-time. }
 \label{fig:teaser}
\end{teaserfigure}

\maketitle

\section{Introduction}
\label{sec:introduction}
Transforming styleless motions to embody a particular character is invaluable for animating characters in feature films or in VR. As the popularity of interactive and immersive VR and AR applications grows, so does the need for real-time user motion characterization. In this paper, we propose a novel technique for transforming a variety of user motions into motions that express a specific character (e.g., Princess, Clown, etc.) in real-time.



We adopt an example-based approach, presuming the availability of a stylistic motion database for the target character. However, the character's body proportions will not be an exact match to the user's, and the character may not even possess a human-like physique. In this work, we posit that a character's distinctive movement style is intricately linked with their unique body shape, so we need to solve both motion stylization and motion retargeting problems jointly. 

Current data-driven motion stylization techniques often perceive style as a consistent feature within data. One approach is to extract style elements from the target examples and apply them to the source motion \cite{aberman2020unpaired,jang2022motion}. Another approach treats style as a conditioning variable for motion generation \cite{park2021diverse,tao2022}. While these methods perform well on stereotypical locomotion datasets, we challenge their assumption by introducing a high quality and diverse dataset of professional performances. Our data clearly illustrates style as context-dependent. For example, the expression of "happiness" in a jumping motion manifests differently than in a crawling motion and their characteristics are not interchangeable. 

To this end, we propose a novel framework that can characterize a user's diverse motions in real-time.
The basic idea is that when user performs a motion, we search for a target character's motion with the most similar context from a motion database and then transfer the style elements of the found motion to the user's motion. However, this approach has the disadvantages of storing character motion database, searching a suitable motion in real-time, as well as limiting target motions to the ones in the database. To improve upon this, inspired by learned motion matching technique~\cite{holden2020learned}, we train a neural network, dubbed Neural Context Matcher (NCM), to generate target motion features suitable for characterizing a source motion instead of searching a database. Modeled as a conditional VAE running autoregessively, the NCM can generate temporally coherent motion features for the target character suitable for the input motion, which is key to obtaining high-quality motion characterization for diverse motions.

In addition, the motions generated by our framework not only reflect the motion style aspects but also the target character's body proportions, making additional motion retargeting to the target character unnecessary. Hence we call our technique \textit{motion characterization}. This is possible because our framework encodes both motion styles and body proportions into motion feature and effectively transfers them to the source motion. To the best of our knowledge, our work is the first that performs motion stylization and retagetting concurrently. Besides, we propose several crucial ideas, such as introducing contrastive loss and incorporating adaptive instance normalization (AdaIN)~\cite{huang2017arbitrary} into transformer decoder, that significantly enhance the motion stylization quality.

Figure~\ref{fig:teaser} shows snapshots of our characterization results. Various motions, such as walking, sitting and jumping (white character), can be characterized to match Zombie, Princess, and Clown. We demonstrate the effectiveness of our framework through comparisons with previous work and ablation study. Additionally, we showcase its capability for real-time live characterization from streamed motion data. We also show that our framework can accommodate sparse inputs, enabling its application in VR tracker-based motion capture systems.  

The major contributions of our work can be summarized as follows:
\begin{itemize}[itemsep=5.0pt]
\item We present the first online motion characterization framework that can transfer both the motion style aspects and body proportions of characters to a variety of user motions.
\item Our C-VAE-based Neural Context Matcher model can effectively generate the target character's motion feature with temporal coherency. 
\item We release a high-quality character motion dataset that contains a total of 6 characters performing various actions, with each action conducted with 5 emotions.
\end{itemize}

\section{Related Work}
\label{sec:related_work}

\subsection{Motion style transfer}
\label{subsec:motion_style_transfer}

Research in motion style transfer centers around two fundamental questions: the definition and representation of style. Conventionally, style is regarded as time-invariant variations within the same motion content, structure, or context, which can vary based on individual characteristics \cite{ma2010modeling} or emotions \cite{Amaya1996Emotion}. A lot of research investigate decomposition algorithms to factorize and parameterize style from motion content \cite{unuma1995fourier, yumer2016spectral, Rose1998Verbs, Shapiro2006Style, Min2009Interactive, Brand2000Style, mason2018few}. 
Style can also be represented non-parametrically or as a label. For instance, Gaussian Process has been used to learn a latent space of pose styles \cite{Grochow2004StyleIK} or motion styles \cite{Wang2007Multifactor, Ikemoto2009Generalizing}. More recently, explicit style labels are used to condition generative models to stylize the output motion \cite{tao2022style, smith2019efficient}. The style labels can also be latent learned from data \cite{park2021diverse}.

Treating style as a static feature limits its representative power. These approaches can often capture stereotypical pose features, but fall short at delivering complex and nuanced characteristics. Xia et al. \shortcite{xia2015realtime} addressed this issue by adapting style parameters in real-time based on local nearest-neighbors. They also contributed a dataset that supported many follow up research. Style may also be spatially varying. Motion Puzzle \cite{jang2022motion} integrates style elements based on body parts into a single character. In our work, we consider style as non-parametric and context-dependent, and therefore choose a different class of approach more akin to motion matching \cite{2016Clavet}.

Recent research in motion style transfer is heavily influenced by the vast body of work in image style transfer. In this context, motion styles are analogous to image textures. The pioneering work of Neural Style Transfer \cite{Gatys2015Neural} introduced the use of the Gram matrix of latent features for style representation, an approach later adopted in the motion domain \cite{holden2016deep, du2019stylistic}. Its successor, the Adaptive Instance Norm (AdaIN) layer \cite{huang2017arbitrary}, has later become the predominant technique with widespread adoption (e.g., \cite{aristidou2022rhythm}). Additionally, Generative Adversarial Networks (GANs), in combination with contrastive and cycle-consistency losses \cite{Zhou2016Learning} have proven effective in self-supervised style learning \cite{dong2017Adult, aberman2020unpaired}, which we also apply in our work.

\subsection{Motion Matching}
\label{subsec:motion_synthesis}
\ignore{
Motion Matching was originally introduced by \cite{2016Clavet}. Created for the game, For Honor, this system was subsequently expanded upon by \cite{2016Clavet}, encapsulating it as a search mechanism across an animation database, aimed at locating the frame that best aligns with the current pose and the user's trajectory. It doesn't necessitate specific labels or contact patterns for the motion data, which enables its application to a myriad of characters, including quadrupeds. The motion matching algorithm can be learned by a neural network, with low memory usage and fast run-time speed~\cite{holden2020learned}.

Inspired by motion matching approach, we train the neural context matcher (NCM) to generate the best matching character feature to input source motion in realtime. .
}

Motion matching \cite{2016Clavet} provides continuous and controllable animation in real-time for interactive gaming via nearest-neighbor search of motion features. To improve its scalability in memory and speed, learned motion matching \cite{holden2020learned} approximates this process using a neural network. Our Neural Contact Matcher (NCM) applies learned motion matching on learned context features to find the closest matching target motion, ensuring real-time and high quality output.

\subsection{Puppeteering and Motion Retargeting}
Puppeteering and motion retargeting can be regarded as a special case of motion style transfer, where the control or source motion must be mapped and adapted to the target character's design and style. Automatically mapping between two arbitrary characters remains a notable challenge with no unique answers. To address this, heuristic-based solution have been developed to analyze either the structure \cite{Kry2009Modal} or motion space between characters \cite{Seol2013Creature, Dontcheva2003Layered}. For the less complex task of mapping between two bipedal skeletons - considered homeomorphic graphs - graph convolutional networks have been successful \cite{aberman2020skeleton, park2021diverse}. We therefore employ this strategy to extra motion features, facilitating retargeting across a diverse range of characters and styles.
\section{Motion data representation and processing}
\label{sec:data_processing}

\begin{figure}[t]
  \begin{minipage}[c]{0.27\linewidth}
    \includegraphics[width=\linewidth]{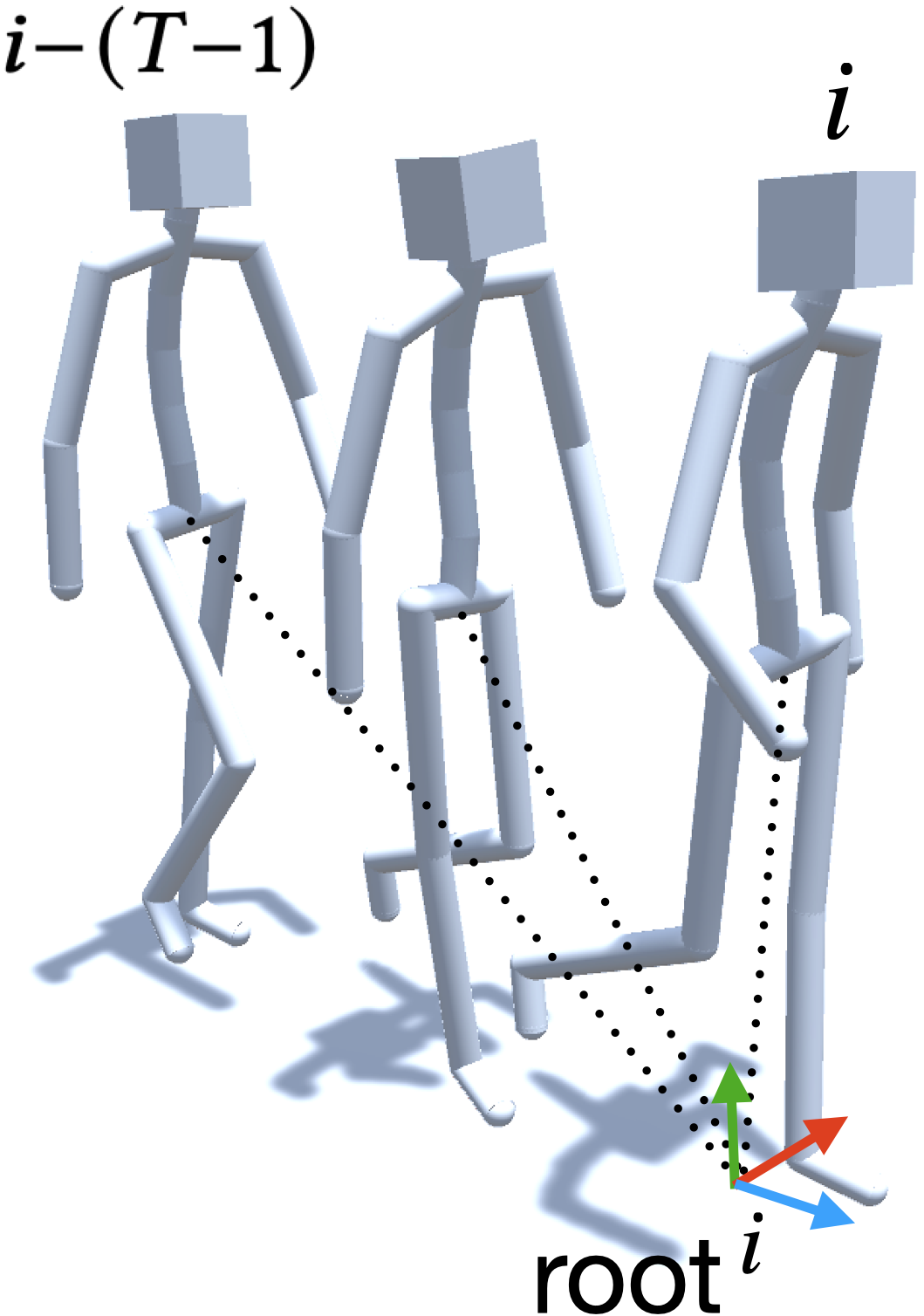}
  \end{minipage}\hfill
  \begin{minipage}[c]{0.6\linewidth}
    \caption{Illustration of the motion representation at frame $i$. 
    We define the character forward facing direction of current frame $i$ as reference frame, which is denoted as $\text{root}^i$.} 
    \label{fig:representation}
  \end{minipage}
\end{figure}

Figure \ref{fig:representation} illustrates the motion representation of our method. The reference frame of a motion, denoted as $\text{root}^{i}$, is located at the ground projection of the pelvis joint at current frame $i$ and aligned to the ground normal direction and the pelvis' forward facing direction.
A motion sequence at frame $i$ is represented in two ways: $\mX^i$ and $\mY^i$ express each element with respect to $\text{root}^{i}$ and with respect to the parent, respectively.
\begin{equation} \label{eq:representation}
    \begin{aligned}
    \mX^i &= [\mathbf{x}]^i_{i-(T-1)}, \,\,\, \mathbf{x} = [x^t_j \, x^r_j \, \dot{x}^t_j \, \dot{x}^r_j],\\
    \mY^i &= [\mathbf{y}]^i_{i-(T-1)}, \,\,\, \mathbf{y} = [y^t_j \, y^r_j \, \dot{y}_j^t \, \dot{y}_j^r],
    \end{aligned}
\end{equation}		
where $j$ denotes all joints (including the pelvis), $x^t \in \real^3$ and $x^r \in \real^6$ (6 for two orthogonal axes) are joint translations and rotations, $\dot{x}^t_j \in \real^3$ and $\dot{x}^r_j \in \real^3$ are joint linear and angular velocities local to $\text{root}^{i}$. 
Likewise, $y^t$, $y^r$, $\dot{y}_j^t$ and $\dot{y}_j^r$ denote joint translation, rotation, and linear and angular velocities local to the parent, with respect to $\text{root}^{i}$. As a result, the total dimensions of our human motion feature with $T$ frames are $\mX \in \real^{T \times n_{joint} \times 15}$ and $\mY \in \real^{T \times n_{joint} \times 15}$, where $n_{joint}$ is the number of joints.

\section{MOCHA Framework}
\label{sec:method}
\begin{figure*}[t]
  \centering
  \includegraphics[width=0.92\textwidth]{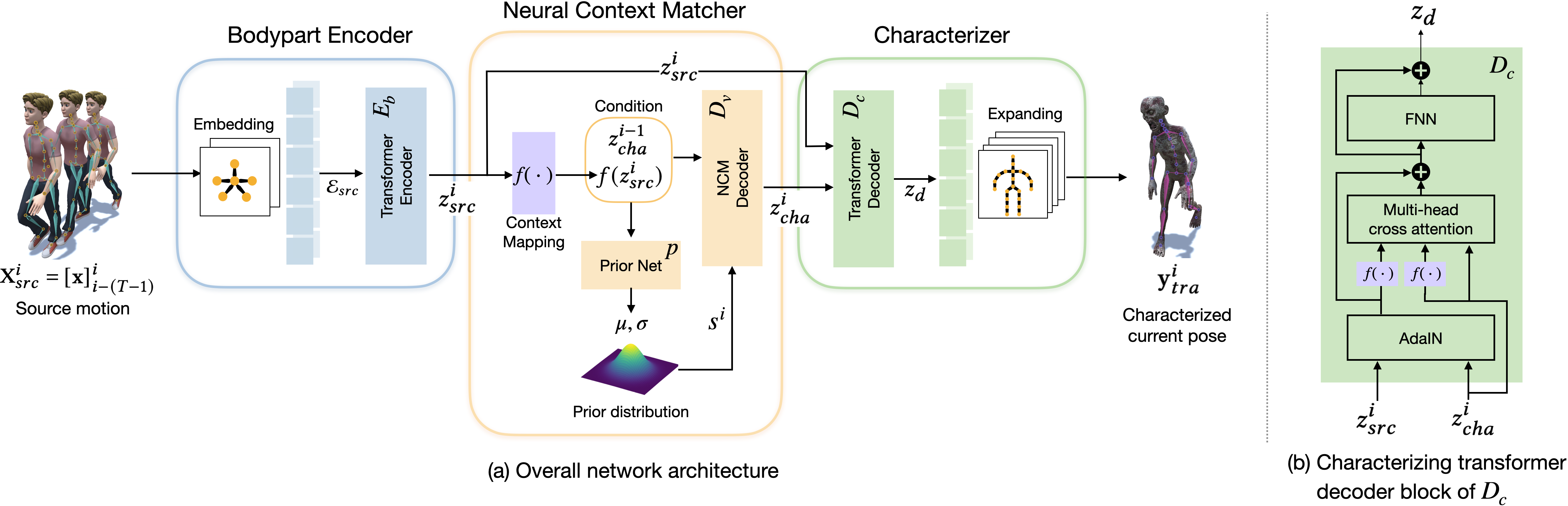}
  \caption{Network configuration. (a) Overall architecture for motion characterization in run-time. Our framework consists of bodypart encoder, neural context matcher, and characterizer networks. (b) Detail of the characterizer transformer decoder block ($D_c$).}
  \label{fig:overview}
\end{figure*}

Figure ~\ref{fig:overview} (a) illustrates our motion characterization framework at runtime. Input to our framework is a motion sequence of length $T$ (1 second, 60 frames) from past to current frame $i$, represented with $\mX^i$.
For output, we generate a characterized motion sequence of the same length, represented as $\mY^i$, but only use the final characterized pose $\mathbf{y}^i$.

Our framework comprises a bodypart encoder, a neural context matcher (NCM), and a characterizer networks. 
At each time frame, the bodypart encoder transforms the input source motion into a feature vector (dubbed \emph{source feature}) that structures human motion into six parts and captures sequential motion dependencies. Subsequently, given a target character, the NCM generates a corresponding feature of the target character (dubbed \emph{character feature}) that shares the most similar context with the source motion. Lastly, the characterizer uses the source feature and the context-matched character feature to synthesize a characterized pose while preserving the source motion's context.

The bodypart encoder and characterizer are character-agnostic, allowing a single trained network to work for all characters. In contrast, the NCM is trained separately for each target character, so the number of NCMs increases linearly with the number of characters. 

\subsection{Bodypart Encoder}
\label{subsec:bodyaprt_encode}
The bodypart encoder in Fig.~\ref{fig:overview} (a) consists of two components: body patch embedding, which reduces the spatial and temporal resolutions of the input motion while preserving bodypart structure, and transformer encoder, which uses a transformer-based structure to learn sequential motion dependencies.

\paragraph{\textbf{Body patch embedding.}} 
Following the approach of flattening 2D patches input for standard vision transformer \cite{dosovitskiy2020image}, we employ body patch embedding, which maintains the graph structure of the human skeleton as much as possible.
Specifically, we use spatial-temporal graph convolutional blocks (STGCN)~\cite{yan2018spatial} to reduce the spatial (joint) and temporal (frame) resolutions. STGCN blocks project an input motion into a sequential feature embedding $\mathcal{E}$. We define embedding process as follows:
\begin{equation} \label{eq:embedding}
    \mathcal{E} = \text{STGCN}(\mX^i).flatten(0,1) + \mathcal{P} \in \real^{(\frac{T}{4}*n_{body}) \times C},
\end{equation}
where $\mathcal{P}$ denotes positional encoding with learnable parameters.
$\mathcal{E}$ comprises a total of $\frac{T}{4}*n_{body}$ patches with $C$ channels, where $n_{body} (= 6)$ is the number of bodyparts (head, spine, arms, and legs). 

\paragraph{\textbf{Transformer encoder.}} We capture spatial-temporal dependencies of body patches by using a transformer structure. We feed the embedding sequence $\mathcal{E}$ into the transformer encoder $E_b$ to generate a source feature $z=E_b(\mathcal{E})$. Each layer of the encoder consists of a multi-head self-attention module (MSA) and a feed-forward network (FFN).
\begin{equation} \label{eq:encoding}
    \begin{aligned}
    z' &= \text{MSA}(\mathcal{E})+\mathcal{E}, \\
    z^i &= \text{FNN}(z')+z' \in \mZ \subset \real^{(\frac{T}{4}*n_{body})\times C}.
    \end{aligned}
\end{equation}

\subsection{Neural Context Matcher}
\label{subsec:context_matching}
After bodypart encoding, we search a character feature that shares a similar context with the source feature. This character feature will later be used to imbue its style aspects into the source feature.
Inspired by learned motion matching (LMM) approach \cite{holden2020learned}, we train the neural context matcher (NCM) to generate the best matching character feature.
Unlike LMM, which performs matching every few frames, NCM runs at every frame to maximize responsiveness to the input motion, which is crucial for producing temporally continuous character features.
To this end, we model the NCM using an autoregressive conditional variational autoencoder. The NCM implicitly models a distribution of possible next character features that match the current source context feature given previous character feature. Samples are drawn from this distribution and passed through the NCM decoder to create a character feature for next frame, one at a time in an autoregressive fashion (yellow part in Fig.~\ref{fig:overview} (a)). 

\paragraph{\textbf{Context mapping}} The context feature $f(z_{src}^i)\in \real^{(\frac{T}{4}*n_{body}) \times C}$ is extracted from encoded feature $z^i$ via context mapping network. The context space made by the context mapping is character-agnostic, capturing shared information on context across character domains. It enables context matching between different characters, which is a crucial step in characterizing.
The context mapping is learned from unlabeled motion data in an supervised manner with a set of loss terms as will be discussed in Sec.~\ref{subsec:context_space}.  

\paragraph{\textbf{Prior Net.}} 
The distribution over possible latent variable $s^i \in \real^C$ for character feature $z_{cha}^{i}$ is described by a learned prior \cite{rempe2021humor} conditioned on previous character feature $z_{cha}^{i-1}$ and current source context feature $f(z_{src}^{i})$:
\begin{equation} \label{eq:prior_net}
    \begin{aligned}
    p(s^i|z_{cha}^{i-1}, \,\, &f(z_{src}^{i})) \\
    &= \mathcal{N}(s^i;\mu(z_{cha}^{i-1}, \, f(z_{src}^{i})), \sigma(z_{cha}^{i-1}, \, f(z_{src}^{i}))),
    \end{aligned}
\end{equation}
which parameterizes a Gaussian distribution with diagonal covariance via a neural network. 

\paragraph{\textbf{NCM decoder.}} 
The character feature $z_{cha}^{i}$ is predicted by the NCM decoder, which takes as input the latent variables $s^i$ while being conditioned on previous character feature $z_{cha}^{i-1}$ and current source context feature $f(z_{src}^{i})$:
\begin{equation} \label{eq:NCM}
    z_{cha}^{i} = D_v(s^i, \, z_{cha}^{i-1}, \, f(z_{src}^{i})).
\end{equation}
We use transformer based C-VAE model for the NCM. 
In training phase, both NCM encoder and decoder are trained as detailed in Section~\ref{subsec:training_stage2} while only the decoder is used for inference. 

\subsection{Characterizer}
\label{subsec:characterizing}
The characterizer transfers the style aspects (e.g., skeleton proportions, characteristic movements, etc.) of the character feature to the source feature. For this, the character transformer decoder with adaptive instance normalization (AdaIN)~\cite{huang2017arbitrary} and multi-head cross-attention generates characterized decoded feature $z_d$ (dubbed \emph{translated feature}), which is then upsampled with De-STGCN blocks to obtain the final characterized motion.

\paragraph{\textbf{Character transformer decoder.}} After context matching, we feed the source feature and character feature into the transformer decoder. The character transformer decoder generates translated feature $z_d=D_c(z_{src}^i,\,z_{cha}^i)$ which merges the context of source motion and the character (style) aspects of target character motion. 

As shown in Fig.~\ref{fig:overview} (b), instead of employing traditional transformer decoder block, we model our decoder with the AdaIN, multi-head cross-attention module (MCA) with context mapping function, and FNN layer.
The AdaIN module transfers the global statistics of character feature, as in ~\cite{jang2022motion}, by taking $z_{src}^i$ as input and injecting character feature $z_{e, cha}^i$ as:
\begin{equation} \label{eq:AdaIN}
    {z_d}'' = 
    \gamma(z_{cha}^i) \left( \frac{z_{src}^i - \mu(z_{src}^i)}{\sigma(z_{src}^i)} \right) + \beta(z_{cha}^i), 
\end{equation}
where $\sigma$ and $\mu$ are the channel-wise mean and variance, respectively. AdaIN scales the normalized $z_{src}^i$ with a learned affine transformation with scales $\gamma$ and biases $\beta$ generated by $z_{cha}^i$.
In the second step, we feed the globally-stylized feature ${z_d}''$ and character feature $z_{cha}^i$ to  MCA. We use the context feature $f({z_d}'')$ to generate the query $Q$, the character context feature $f(z_{cha}^i)$ to generate the key $K$, and the character feature $z_{cha}^i$ to generate the value $V$:
\begin{equation} \label{eq:QKV}
    Q = f({z_d}'')W_q, \,\,\,\, K = f(z^i_{cha})W_k, \,\,\,\, V = z^i_{cha}W_v,
\end{equation}
where $W_q, W_k, W_v \in \real^{C \times d_{head}}$.
Then, the output sequence $z_d$ of the transformer decoder is obtained by
\begin{equation} \label{eq:decoding}
    \begin{aligned}
    z'_d &= \text{MCA}(Q, K, V)+{z_d}'', \\
    z_{d} &= \text{FNN}(z'_d)+z'_d.
    \end{aligned}
\end{equation}

\paragraph{\textbf{Body patch expanding and output}}
The translated feature $z_d \in \real^{(\frac{T}{4}*n_{body}) \times C}$ is upsampled with additional De-STGCN blocks, which has symmetric architectures to that of STGCN, to yield output translated motion $\mY_{tra}^i$ as follows:
\begin{equation} \label{eq:expanding}
    \mY_{tra}^i = \text{De-STGCN}(Reshape(z_d)). 
\end{equation}
Finally, we pick the last frame pose $\mathbf{y}^{i}_{tra}$ from $\mY_{tra}$ as final output.

\paragraph{\textbf{Root motion}}
We maintain the root angular velocity of the source motion while scaling the linear velocity according to the ratio of the average hip velocities of the source motion $\mX^i$ and the output motion $\mY_{tra}^i$. This strategy allows for maintaining the overall shape of the root trajectory of the source motion, while varying the global linear velocity to match the target character.
\section{Training}
\label{sec:training}
Our training pipeline consists of two-stages; the bodypart encoder and characterizer are trained first, followed by training the NCM.
Thus, in the first stage, our framework learns to extract encoded features from a motion and transfer style aspects of one motion (e.g., motion style and skeleton proportions) to the other motion to synthesize a characterized motion. At this stage, the context mapping network is also trained as a part of transformer decoder. The first stage training is conducted in an unsupervised way.
The second stage trains the NCM in a supervised manner to generate target character features corresponding to the input context features from the source. 

\subsection{Stage-1}
\label{subsec:training_stage1}
\begin{figure}[t]
  \centering
  \includegraphics[width=0.95\linewidth]{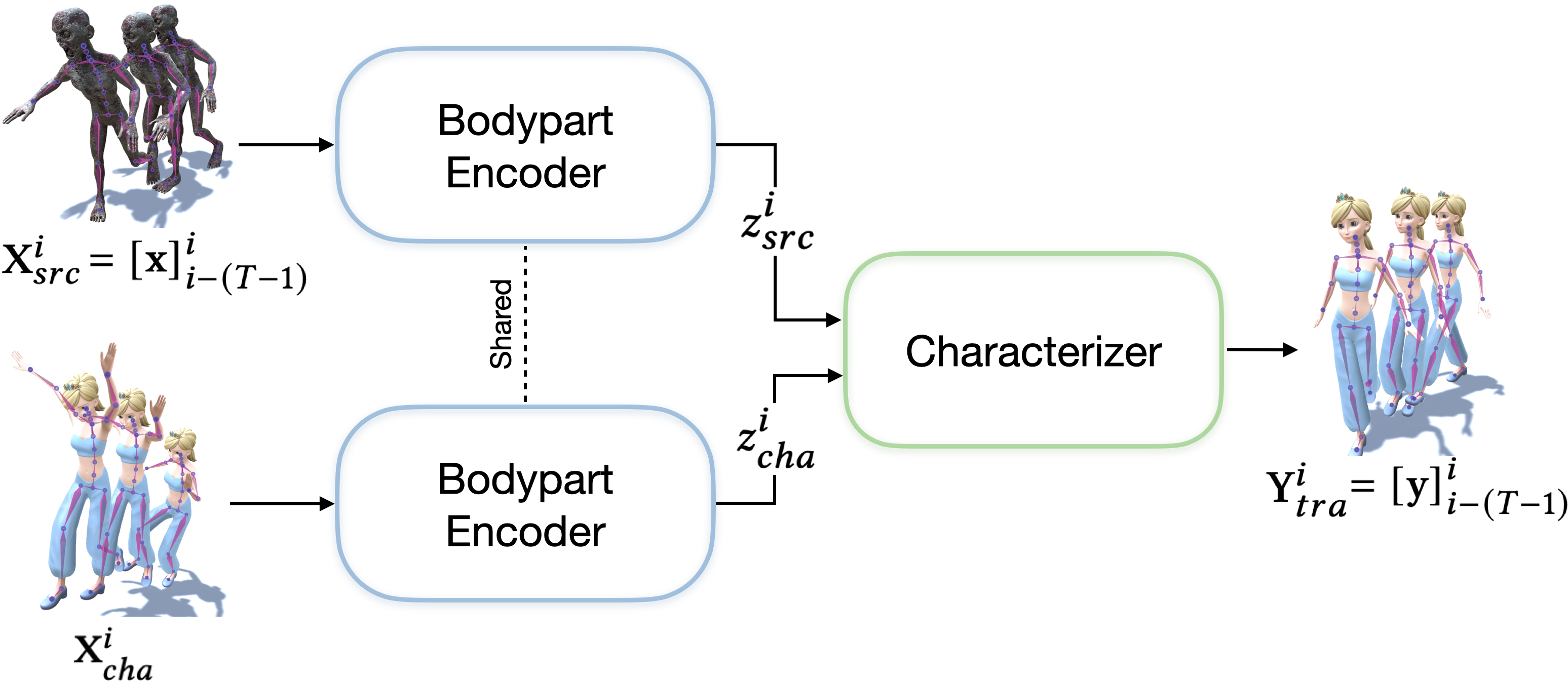}
  \caption{Stage-1 training process. A source motion $\mX_{src}^i$ and a target motion $\mX_{cha}^i$ are randomly selected from different characters in the motion dataset, and the networks are trained to make the characterized motion $\mY_{tra}^i$ preserve the context of $\mX_{src}^i$ while reflecting the characteristic aspects of $\mX_{cha}^i$.}
  \label{fig:training_chz}
\end{figure}

Figure~\ref{fig:training_chz} shows the training architecture in stage-1.
Without the NCM, we randomly choose a source motion $\mX_{src}^i$ and a character motion $\mX_{cha}^i$ to train the bodypart encoder and the characterizer jointly. In addition to using common loss terms for learning style transfer, such as identity and cyclic consistency losses, we newly introduce a contrastive loss to enhance local context preservation both spatially and temporally.
To simplify the notation, we use $BE(\cdot)$ and $CH(z_{src},z_{cha})$ to denote body part encoder and characterizer, and omit the superscript $i$. 

\paragraph{\textbf{Identity and cycle loss}}
To design identity and cycle consistency losses, we first define a reconstruction loss that computes the difference of two motions both in terms of $\mX$ and $\mY$~\cite{holden2020learned} as well as their velocities. 
\begin{equation*} \label{eq:rec_total}
    \begin{aligned}
    \mathcal{L}_{rec}&(\widetilde{\mX}, \, \mX) = \lambda_{loc} \mathbb{E}_{\mX}[\| \widetilde{\mX}- \mX\|_1]+ \lambda_{rt} \mathbb{E}_{\mY}[\| \widetilde{\mY}- \mY\|_1] \\
    &+ \lambda_{lvel}\mathbb{E}_{\mX}[\| V(\widetilde{\mX})- V(\mX)\|_1] + \lambda_{rvel}\mathbb{E}_{\mY}[\| V(\widetilde{\mY})- V(\mY)\|_1],
    \end{aligned}
\end{equation*}
where $\mX=FK(\mY)$, $V(\mX) = \frac{\mX^0 - \mX^1}{h}, \; V(\mY) = \frac{\mY^0 - \mY^1}{h}$, $h$ is time step, and $\lambda_{loc}$, $\lambda_{rt}$, $\lambda_{lvel}$, $\lambda_{rvel}$ are the relative weights. $V(\mathbf{X})$ represents the rate of change of $\mathbf{X}$. Part of $V(\mX)$ corresponds to the joint accelerations.

The identity loss ensures that the input motion remains unchanged when it is used for both the source and character motions: 
\begin{equation} \label{eq:identity}
\mathcal{L}_{id} = \mathcal{L}_{rec}(\mX^{id}_{src}, \, \mX_{src}) + \mathcal{L}_{rec}(\mX^{id}_{cha}, \, \mX_{cha}), 
\end{equation}
where $\mX^{id}_{src} = FK(CH(z_{src},\, z_{src}))$ and $z_{src}=BE(\mX_{src})$. $\mX^{id}_{src}$ is obtained by feeding $\mathbf{X}_{src}$ to both $\mathbf{X}_{src}$ and $\mathbf{X}_{cha}$ in Fig. \ref{fig:training_chz}. Likewise, $\mX^{id}_{cha} = FK(CH(z_{cha},\, z_{cha}))$ with $z_{cha}=BE(\mX_{cha})$.

To guarantee that the resulting motion $\mY_{tra}$ (hence $\mX_{tra}$) preserves the context of the source motion $\mX_{src}$ and the characteristics of the character motion $\mX_{cha}$, we employ cycle consistency loss \cite{choi2020stargan}.
\begin{equation} \label{eq:cyc}
    \mathcal{L}_{cyc} = \mathcal{L}_{rec}(\mX^{cyc}_{src}, \, \mX_{src}) + \mathcal{L}_{rec}(\mX^{cyc}_{cha}, \, \mX_{cha}), 
\end{equation}
where $\mX^{cyc}_{src} = FK(CH(z_{tra}, \, z_{src}))$, $\mX^{cyc}_{cha} = FK(CH(z_{cha}, \, z_{tra}))$, $z_{tra} = BE(FK(\mY_{tra}))$ and $\mY_{tra}=CH(z_{src},\, z_{cha})$. $\mathbf{X}^{cyc}_{src}$ is obtained by feeding $\mathbf{X}_{src}$ and $\mathbf{X}_{cha}$ to top and bottom inputs in Fig 4 to get $\mathbf{Y}_{tra}$, followed by feeding $\mathbf{X}_{tra} (= FK( \mathbf{Y}_{tra}))$ and $\mathbf{X}_{src}$ to top and bottom inputs in Fig. \ref{fig:training_chz} to get $\mathbf{Y}^{cyc}_{src}$.

\begin{figure}[t]
  \centering
  \includegraphics[width=0.8\linewidth]{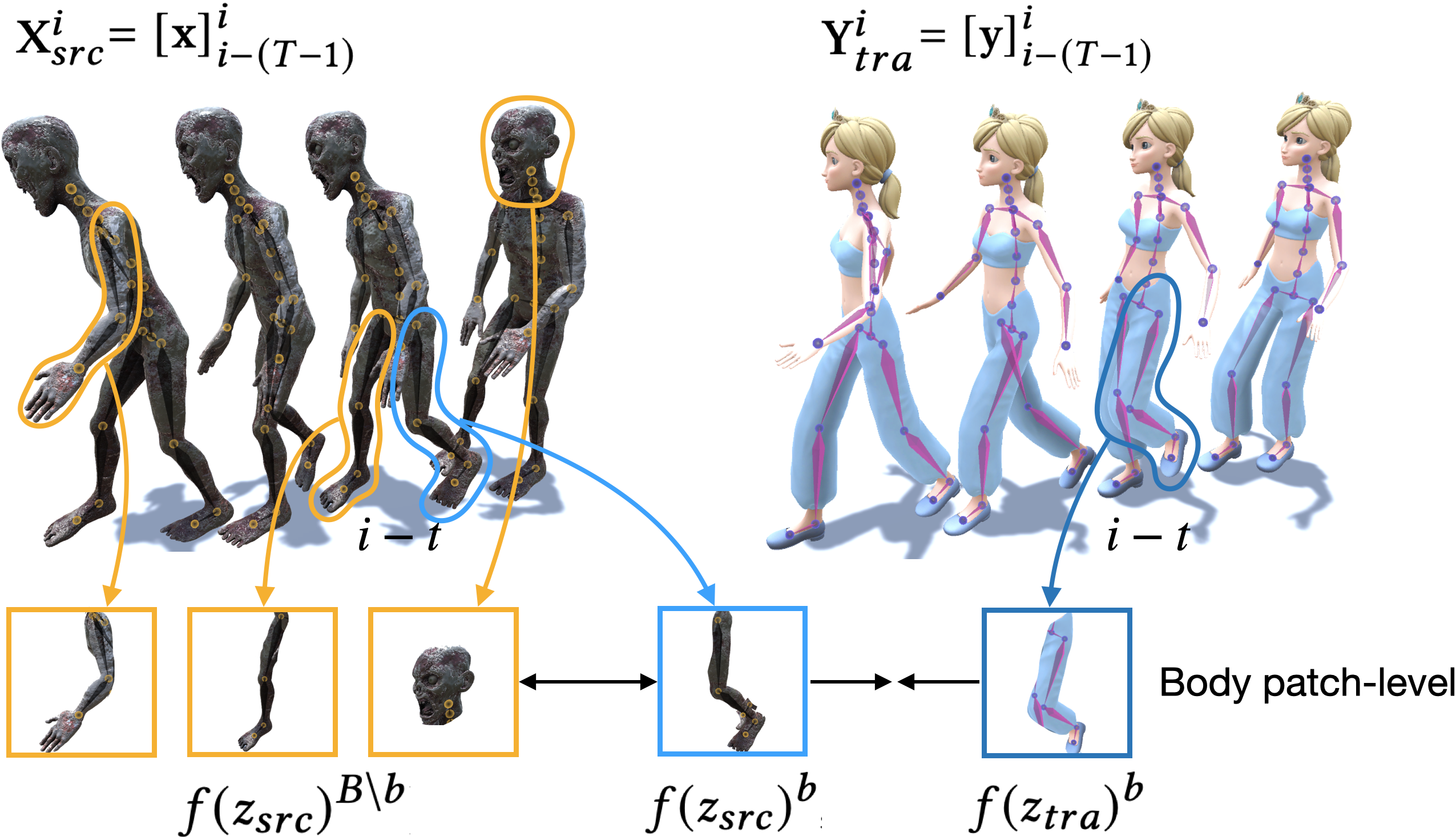}
  \caption{Illustration of body patch-level context contrastive loss. Blue box indicates a positive sample, while yellow boxes denote negative samples.}
  \label{fig:contrastive}
\end{figure}

\paragraph{\textbf{Body patch contrastive loss}}  
To ensure that the characterized output motion not only maintains the overall context $f(z_{src})\in \real^{(\frac{T}{4}*n_{body}) \times C}$ of source motion $\mX_{src}$, but also preserves context of body patches at a specific location between source and output, we introduce a body patch contrastive loss. For example, in Figure \ref{fig:contrastive}, the context of a princess leg at frame $i-t$ should be closer to that of input zombie leg at frame $i-t$ than the other patches of the same input.

To define the body patch-level context loss between $f(z_{src})$ and $f(z_{tra})$, we use infoNCE loss~\cite{oord2018representation}: 
\begin{equation} \label{eq:infoNCE}
    l(\hat{v},v^\text{+},v^\text{-} ) = -\text{log}\left[ \frac{\text{exp}(\hat{v} \cdot v^\text{+}/\tau)}{\text{exp}(\hat{v} \cdot v^\text{+}/\tau) + \sum_{n=1}^N \text{exp}(\hat{v} \cdot v^\text{-}_n/\tau)} \right],
\end{equation}
where $\tau$ is the temperature parameter, and $v^\text{+}$ and $v^\text{-}$ denote positive and negative for $\hat{v}$.

We set pseudo positive samples between body patch-level context of source  $f(z_{src})$ and characterized motion $f(z_{tra})$; for a body patch $f(z_{tra})^b \in f(z_{tra})$, we set its positive patch $f(z_{src})^b$ as the patch in the same location in $f(z_{src})$, and negative patches $f(z_{src})^{B \setminus b}$ as all other patches.
\begin{equation} \label{eq:contrastive}
    \mathcal{L}_{ctr} = \mathbb{E}_{\mX}\sum_{b}^{B}l(f(z_{tra})^b, f(z_{src})^b, f(z_{src})^{B \setminus b}),
\end{equation}
where $b \in \{1,2, \dots B\}$ and $B \,(={T}/{4}*n_{body})$ is the number of body patches. Figure~\ref{fig:contrastive} illustrates how body patch-level contrastive context loss work, with procedure to define positive and negative samples. 

The total loss function of stage-1 is thus:
\begin{equation} \label{eq:stage_1}
    \mathcal{L}_{stage1} = \lambda_{id}\mathcal{L}_{id} + \lambda_{cyc}\mathcal{L}_{cyc} + \lambda_{ctr}\mathcal{L}_{ctr},
\end{equation}
where $\lambda_{id}$, $\lambda_{cyc}$, and $\lambda_{ctr}$ are weights.

\subsection{Stage-2}
\label{subsec:training_stage2}
\begin{figure}[t]
  \centering
  \includegraphics[width=0.8\linewidth]{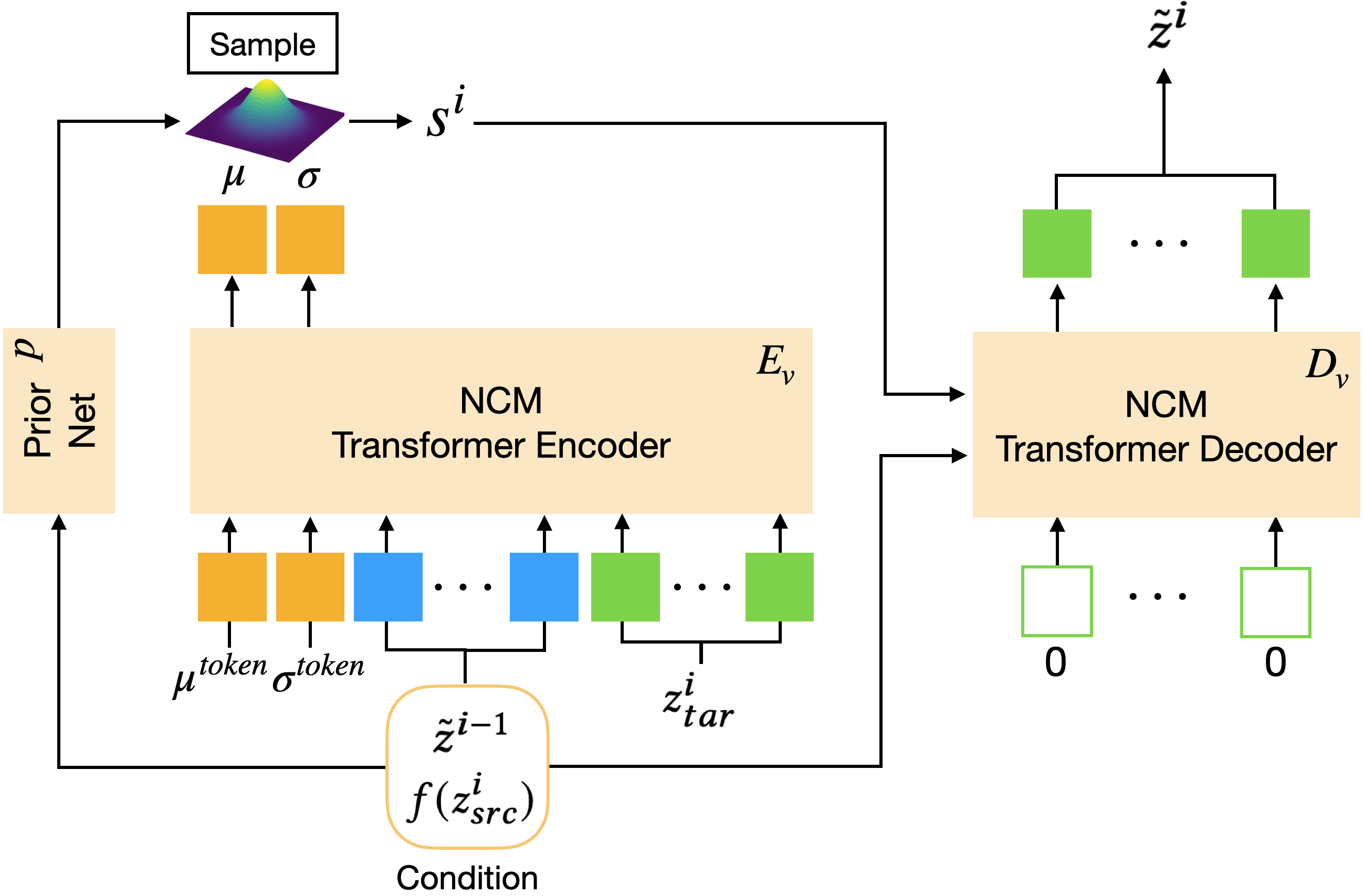}
  \caption{Training architecture of Neural Context Matcher. Once trained, only the decoder is used for inference.}
  \label{fig:training_NCM}
\end{figure}

In stage-2, the NCM is trained to infer a target character feature $z_{cha}$ that shares the same context with a source feature $z_{src}$. Briefly, this is achieved by providing the ground truth character feature $z_{cha}$ obtained from searching a target feature database $\mathcal{D}_{tar}=[\mZ_{tar},f(\mZ_{tar})]$ with a context metric $||f(z_{tar})-f(z_{src})||$.
To begin with, we roughly annotate the motion data with action labels (e.g., walk, jump, and crawl) to facilitate the nearest neighbor search; limiting the search only on the relevant subset of the same action label saves time and enhances accuracy of the results. 

As the NCM operates autoregressively, training the NCM encoder and decoder is conducted with a sequence of source features $z_{src}^i|_{i=0}^s$ and its corresponding sequence of target features $z_{cha}^i|_{i=0}^s$ obtained with nearest neighbor search. These feature sequences represent continuous motions in motion database.
In the training, the NCM transformer takes the previous character feature and current source context feature as condition to reconstruct the context-matched target character feature. At the same time, it attempts to shape the latent variable $s$ as a standard normal distribution ($\mu_{pr}$, $\sigma_{pr}$) using $PriorNet$. This process is illustrated in Fig.~\ref{fig:training_NCM}. 
For details on the training procedure, please see Algorithm 1 in supplementary material.

At run-time, the encoder is discarded and the decoder is used to predict matched character features. At each time frame, we pass $f(z_{src}^i)$ through decoder to predict a matched character feature $z_{cha}^{i}$ (Sec.~\ref{subsec:context_matching}).

\section{Evaluation and Experiments}
\label{sec:Experiments}

We conduct comparisons with other methods, perform an ablation study, and conduct additional experiments to demonstrate the utility of our framework. 
For visual animation results, please refer to the supplementary video.

\subsection{Datasets}
\label{subsec:dataset}
We constructed a high-quality character motion dataset with five professional actors. 
The dataset comprises a total of 6 characters (Clown, Ogre, Princess, Robot, Zombie, and AverageJoe) performing various actions (Dance, Fight, Jump, Crawling, Run, Walk, and Sit) with 5 emotion variations for each action (Angry, Happy, Neutral, Sad, and Scared). 
Every character preserves the same body proportions as the actors, except we scale the forearm of the Ogre character to be twice as long to convey its style. 
The dataset contains a total of 573k frames, captured at 60 fps, resulting in approximately 159 minutes of data. 
Since different emotions are manifested with unique styles, we consider a character-emotion pair as an individual character, such as ``Neutral Zombie'' and ``Happy Ogre''.

We additionally tested our algorithm on the Adult2child dataset \cite{Dong2020Adult} for quantitative evaluation.
It consists of 17 subjects including nine adults (older than 18 years) and eight children (5-10 years old) performing actions such as Jump as high as you can in place, Punch, Kick, Walk, and Hop Scotch. We treat each subject as a character due to their personalized styles. Adult2child dataset is suitable for our evaluation since the skeleton proportions and motion styles vary depending on the subject.


\subsection{Qualitative evaluation}
\label{subsec:qualitative}
First, we qualitatively compare our method with two baselines: Motion Puzzle~\cite{jang2022motion} and replacing NCM in MOCHA with nearest-neighbor search (Nearest Neighbor).
Since Motion Puzzle only works for a single body proportion, we have to replace its reconstruction loss with ours to accommodate skeleton variations at training. Note that Motion Puzzle conducts an offline, full sequence-to-sequence translation, while MOCHA is an online translator.

\begin{figure*}[t]
  \centering
  \includegraphics[width=0.99\textwidth]{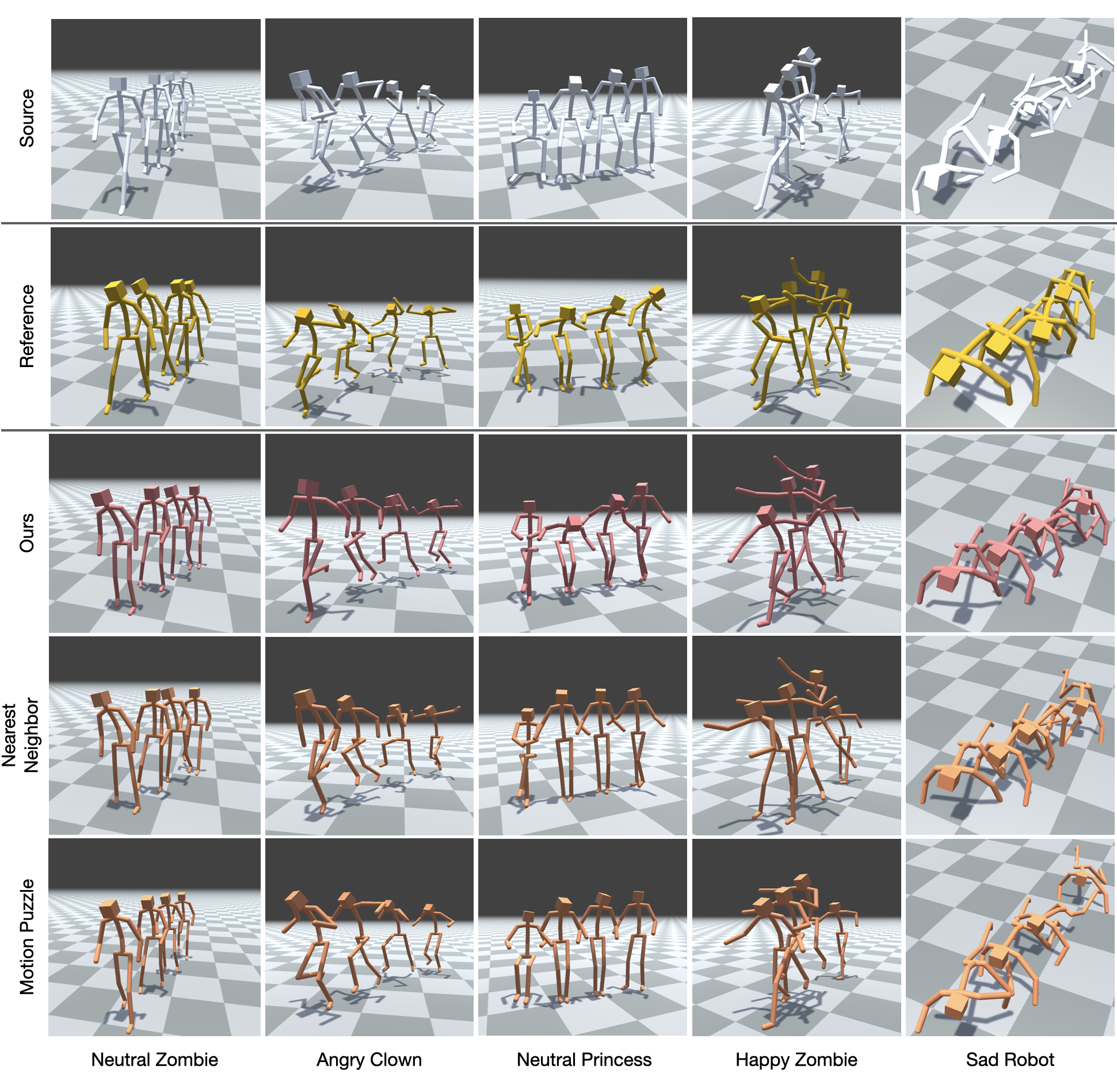}
  \caption{Qualitative evaluation. Source motions of Neutral AverageJoe (top) are characterized with each method (3-5 rows). Reference motion is used for the Nearest Neighbor and Motion Puzzle.
  \newline
  \newline
  \newline
  \newline
  \newline
  \newline
  \newline
  }
  \label{fig:qualitative}
\end{figure*}

Figure~\ref{fig:qualitative} compares the characterization result of Neutral AverageJoe as the source (1st row) among MOCHA, Nearest Neighbor, and Motion Puzzle. Unlike MOCHA, both Motion Puzzle and the Nearest Neighbor method require a specific reference motion as input. Therefore, we manually selected a reference from each target character performing the most similar action (2nd row) for them. 

Compared to Motion Puzzle, MOCHA better preserves the unique style in each target character. Our results accurately capture the limping legs of the Neutral Zombie, the energetic waves of the Angry Clown, the elegantly crossed legs of the Princess, and the wildly flailing arms of the Happy Zombie. In contrast, Motion Puzzle dampens the target style in favor of the source motion. 


Nearest Neighbor exhibits noticeable discontinuity in the output, due to the relatively sparse context space and insufficient temporal features used in search. It cannot respect the source motion context when similar context is not available for the target character. Thanks to the autoregressive CVAE, our method using NCM can mitigate these issues to produce smooth and consistent output.



\subsection{Quantitative evaluation}
\label{subsec:quantitative}
We quantitatively compare the degree of motion quality, context preservation, and style reflection with Motion Puzzle and the Nearest Neighbor method. Following prior work \cite{park2021diverse,jang2022motion}, we use Frechet Motion Distance (FMD), context recognition accuracy (CRA), and style recognition accuracy (SRA) as metrics. We train the CRA classifier and the SRA classifier using vision transformer~\cite{dosovitskiy2020image} for its superior accuracy. 
The two datasets, MOCHA and Adult2child, are each split into a 90$\%$ training set and a 10$\%$ test set. Results are detailed in Table~\ref{tab:qulitative}. We didn't apply CRA on the MOCHA dataset because it does not contain action labels. We also have to remove Motion Puzzle from the Adult2Child dataset comparison as it fails to train. 
When compared to Motion Puzzle, our method yields superior FMD and SRA scores, demonstrating higher motion quality and a more precise style reflection. A comparison with the Nearest Neighbor method underscores the significance of our NCM module, as we consistently achieve higher SRA scores due to superior temporal coherence. The better CRA score for Nearest Neighbor can be attributed to its direct usage of features from the database.
\ignore{

Comparing with Motion Puzzle on MOCHA dataset, our method shows better FMD and SRA scores, which suggests that our method produces more plausible motions that reflect target styles better than Motion Puzzle.
As the only difference of our method and the Nearest Neighbor is the existence of NCM, comparison with the Nearest Neighbor reveals the benefit of the NCM. For both datasets, our method shows consistently higher SRA, suggesting that temporally coherent target features generated by the NCM improves the style reflection quality. On the other hand, the Nearest Neighbor shows higher CRA, which is somewhat reasonable because it searched target features directly from the database. 
}


\begin{table}[t]
\small
\begin{subtable}[h]{0.45\textwidth}
\begin{tabular}{ccccc}
\hline
\multirow{2}{*}{Methods} & \multirow{2}{*}{FMD$\downarrow$} & \multicolumn{2}{c}{SRA$\uparrow(\%)$}         & \multicolumn{1}{c}{CRA$\uparrow(\%)$} \\
                         &                      & Top1           & Top5           &                   \\ \hline
Ground truth             &                      & 43.61          & 84.76          & -                    \\ \hline
Ours                     & {\underline{34.99}}          & \textbf{36.51} & \textbf{73.68} & -                    \\
Nearest Neighbor         & \textbf{32.80}       & 23.62          & 69.06          & -                    \\
Motion Puzzle            & 39.51                & {\underline{31.78}}    & {\underline{70.7}}     & -                    \\ \hline
\end{tabular}
\caption{MOCHA dataset.}
\label{tab:eval_MOCHA}
\end{subtable}

\begin{subtable}[h]{0.45\textwidth}
\begin{tabular}{cccccc}
\hline
\multirow{2}{*}{Methods} & \multirow{2}{*}{FMD$\downarrow$} & \multicolumn{2}{c}{SRA$\uparrow(\%)$} & \multicolumn{2}{c}{CRA$\uparrow(\%)$} \\
                         &                                  & Top1              & Top5              & Top1              & Top5              \\ \hline
Ground truth             &                                  & 96.46             & 99.77             & 58.97             & 85.60             \\ \hline
Ours                     & \textbf{25.98}                   & \textbf{68.57}    & \textbf{95.21}    & 38.18             & 74.45             \\
Nearest Neighbor         & 29.30                            & 63.32             & 93.20             & \textbf{45.96}    & \textbf{79.55}    \\
\hline
\end{tabular}
\caption{Adult2child dataset.}
\label{tab:eval_adult2child}
\end{subtable}


\caption{Frechet Motion Distance (FMD), context recognition accuracy (CRA), and style recognition accuracy (SRA) on samples generated by each method as quantitative comparison.}
\label{tab:qulitative}
\end{table}

\subsection{Ablation study}
\label{subsec:ablation}
\begin{figure*}
  \centering
  \includegraphics[width=0.5\linewidth]{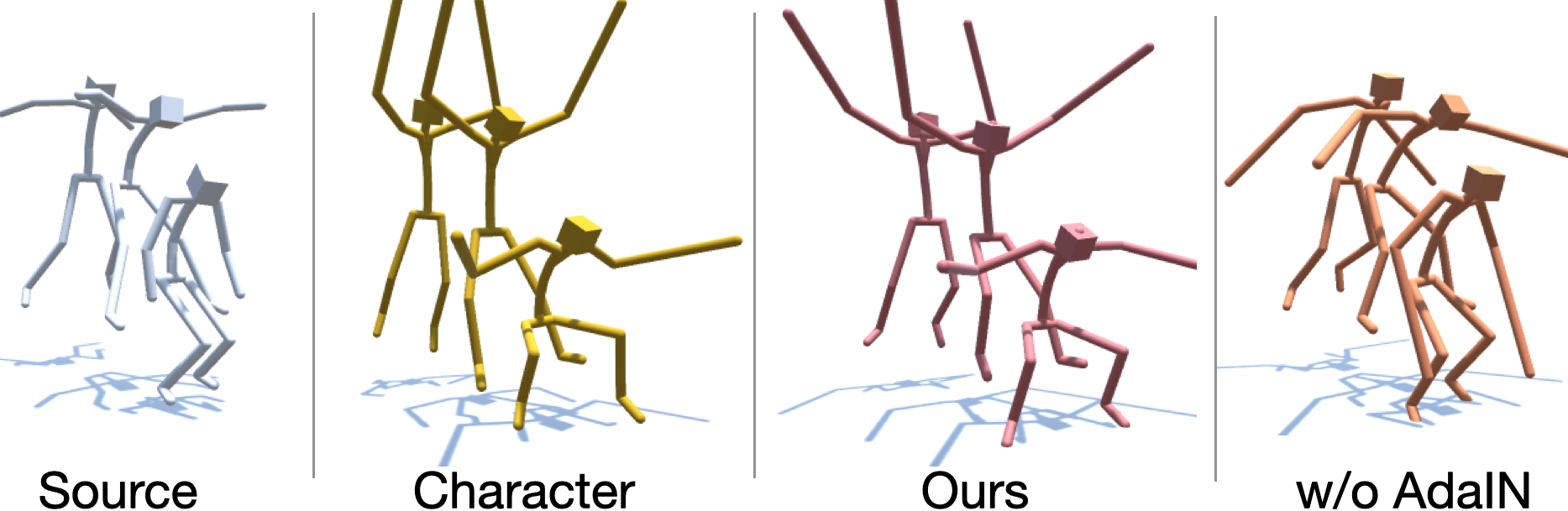}
  \caption{Ablation study on AdaIN. Result obtained without AdaIN fails to capture the jumping style of Ogre.}
  \label{fig:ablation_AdaIN}
\end{figure*}

We conducted ablation studies to demonstrate the effectiveness of the AdaIN layer and the contrastive loss. For additional ablation studies, please refer to the supplementary material.
\paragraph{\textbf{Effect of AdaIN}} 
The AdaIN module facilitates successful characterization by incorporating a global style before the cross attention layer in the characterizing transformer decoder block (Figure~\ref{fig:overview} (b)). To assess its impact, we trained a Characterizer without the AdaIN module. 
Results in Figure~\ref{fig:ablation_AdaIN} shows that the ablated model cannot capture the Ogre's raising arms style without using AdaIN. 

\begin{figure*}
  \centering
  \includegraphics[width=0.5\linewidth]{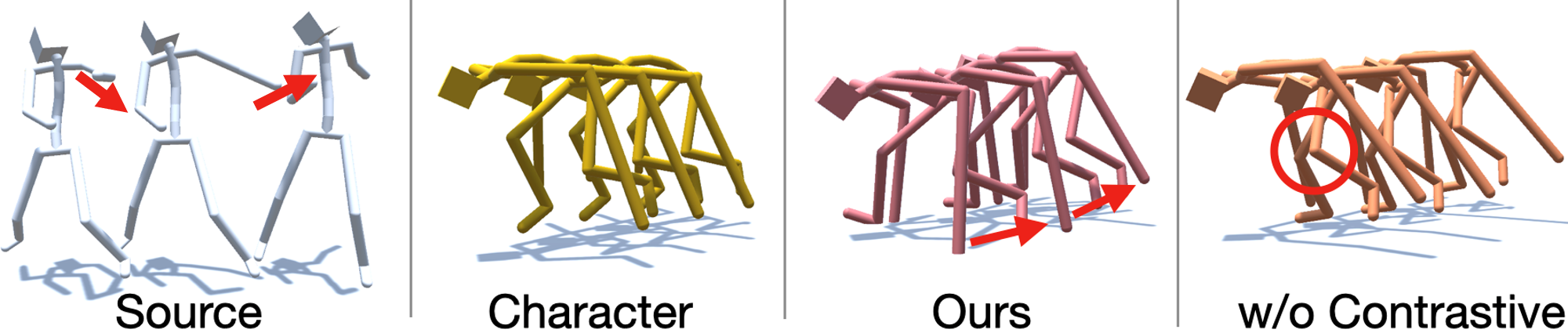}
  \caption{Contrastive loss test. Ogre's left and right leg are mistakenly swapped (red circle) without the contrastive loss. Furthermore, the contrastive loss helps retain the left arm's swing phase of the source motion to the characterized motion (red arrows).}
  \label{fig:ablation_ctr}
\end{figure*}

\paragraph{\textbf{Effect of contrastive loss}}
To examine the effect of patch-level contrastive loss, we tested stage-1 training without the contrastive loss. Since the advantage of our contrastive loss is to preserve spatial-temporal context features across all characters, we randomly picked a source and a character motion to generate output.
Figure~\ref{fig:ablation_ctr} shows that the left and right legs of the output Ogre character are swapped (red circle) without patch-wise contrastive loss because attention module alone may mismatch the left leg of the source with the right leg of the character. In contrast, our model can distinguish the left and right side of the body. Additionally, despite significant differences of two motions, our method reasonably preserves the arm swinging motion of the source (red arrow), which cannot be achieved with the ablation model.

\subsection{Context space analysis}
\label{subsec:context_space}

\begin{figure*}
  \centering
  \includegraphics[width=0.45\linewidth]{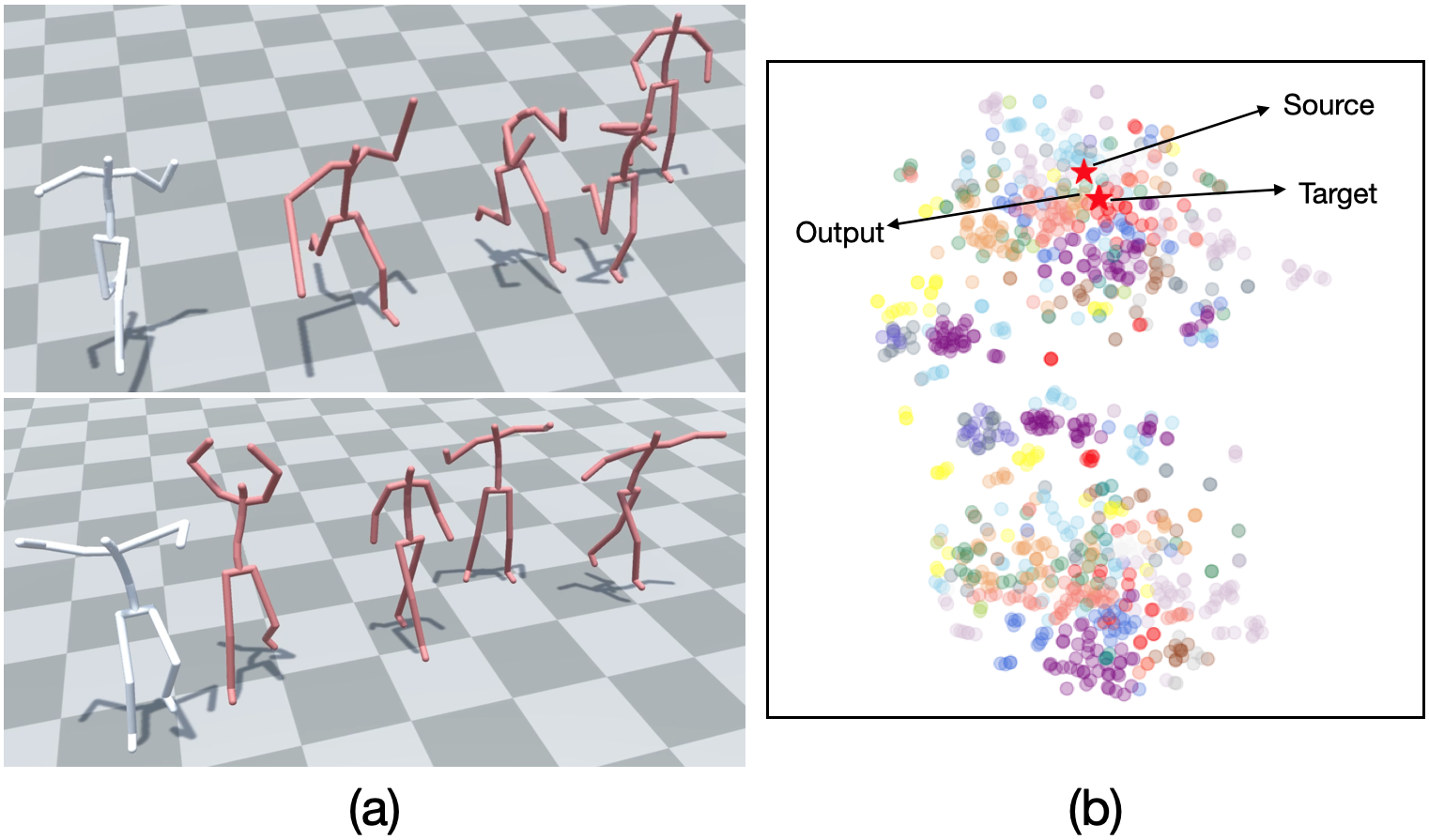}
  \caption{(a) Four best matching motions in the entire datasets given the white character's motion (top: running, bottom: walking).  (b) A t-SNE visualization of the entire context features in the test dataset with colors differentiating characters. Three context feature points corresponding to a source, matched target, and output motion are located very closely to each other.}
  \label{fig:context}
\end{figure*}

Successful context matching across different characters is possible thanks to the shared character-agnostic context space constructed by the context mapping module. To visualize its result, we randomly sampled a character motion and computed its context feature as a query, then searched another character's context database for the closest match. Figure~\ref{fig:context} (a) shows semantically similar yet diverse postures across four different characters that match the query.  
In another experiment, given a source motion ($\mX_{src}$), we searched for the nearest context feature ($z_{cha}$) of a target character and generated a characterized motion ($\mY_{tra}$). Subsequently, both source and characterized motions were mapped to the context space as shown in Figure~\ref{fig:context} (b). One can see that all three context feature points ($f(BE(\mX_{src})), f(z_{cha}), f(BE(\mY_{tra}))$)  
are located very close together, showing that the learned motion context indeed shares the same space across characters.

\subsection{Applications}
\label{subsec:applications}

\begin{figure*}
  \centering
  \includegraphics[width=0.5\linewidth]{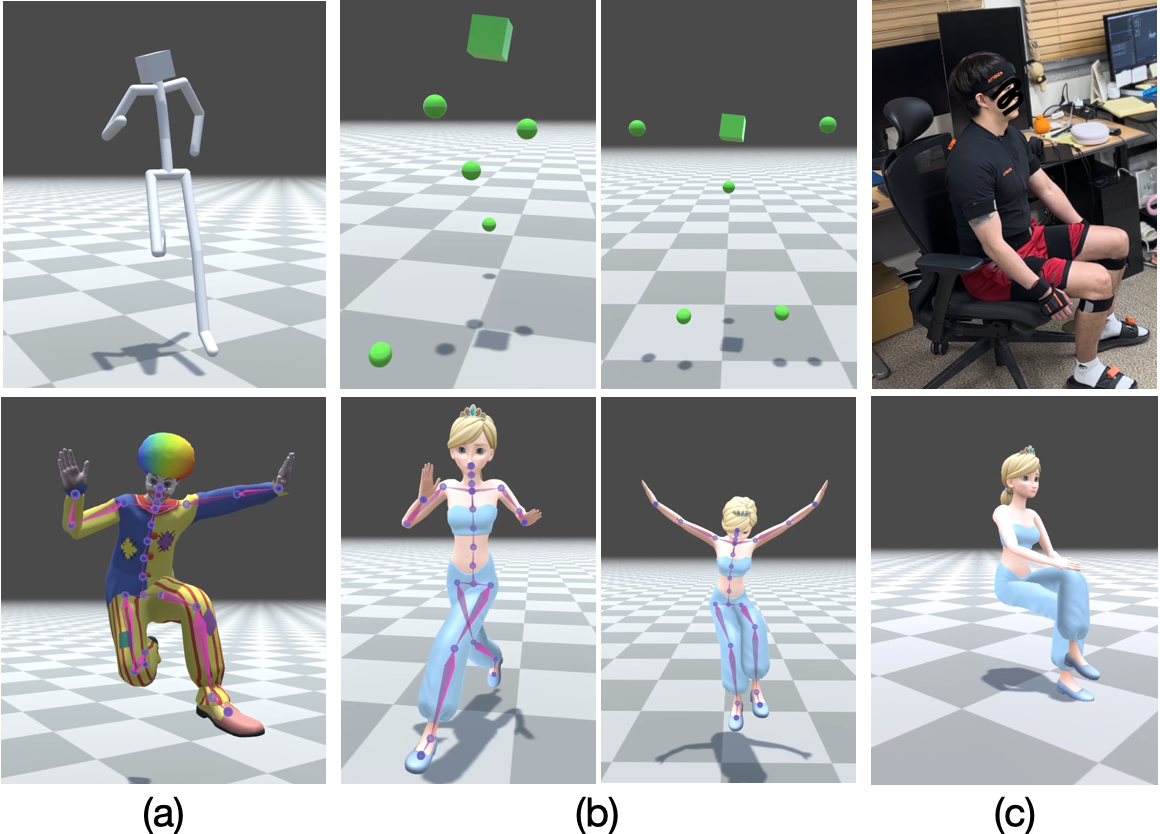}
  \caption{Applications of our framework. Characterization results from an unseen subject (left), with sparse input from 6 joints (middle), and from streamed motion data (right).}
  \label{fig:applications}
\end{figure*}

\paragraph{\textbf{Input from unseen subjects}}
We test the effectiveness of our framework for inputs from unseen subjects with varying heights. Figure~\ref{fig:applications} (a) shows that our framework works successfully even when motions from unseen subjects with different body proportions are given as input.


\paragraph{\textbf{Sparse input}}
\label{subsec:sparse}
We supposed that our framework could work with sparse inputs from the hip and end-effectors as they may contain essential information about context and style. To verify this, we reduced the input joints to only six (hip, head, hands, and feet) and retrained the entire framework. As shown in Figure~\ref{fig:applications} (b), our framework successfully characterizes motion as Neutral Princess from sparse input. 
This experiment suggests that our framework can accommodate 6-point tracker input data for real-time applications.

\paragraph{\textbf{Live characterization from streamed motion data}}
\label{subsec:real-time}
We demonstrate the ability of our framework to characterize streamed motion data in real-time. Figure \ref{fig:applications} (c) shows a snapshot of live characterization of a streamed motion captured with Xsens Awinda sensor.
The supplementary video shows that our method can produce successfully characterized motion even with network delays and noisy input. The overall inference time is under 16ms, achieving 60Hz or higher frame rate with a 2080 Ti GPU. 

\section{Discussion and Conclusion}
\label{sec:limitation}

We introduced MOCHA, a motion characterization framework that can transform user motions to embody distinct style of characters in real-time, and demonstrated its effectiveness through a number of experiments and analysis.

Our method has several limitations that necessitates further exploration.
First, our method is most effective when a character motion dataset contains a single characteristic style per context. If multiple motions with different styles share the same context for a target character, our NCM may encounter difficulties in generating temporally consistent character features, leading to discontinuous motions. We observed such phenomena when consolidating all emotion sets of a character into one, thereby allowing emotional style variations within the same motion. Future work is needed to enable such style variations even within a single character.

Second, while our unsupervised learning of context mapping is effective without manual style labels, it is not flexible enough to encompass all stylistic diversity of the character data.
For instance, Angry Ogre motions in our database has a unique style of running on all fours and jumping sideways. Our algorihtm currently could not link them to the running and jumping motions of other characters, as shown in the supplementary video. More advanced methods for context learning is desirable to improve the characterization quality.

\begin{acks}
This work was conducted during Deok-Kyeong Jang's internship at Meta. Deok-Kyeong Jang would like to thank Michelle Hill for her help in capturing the dataset and the Meta team for their mentorship and support during his internship. Jungdam Won was partially supported by the New Faculty Startup Fund from Seoul National University, ICT (Institute of Computer Technology) at Seoul National University. Sung-Hee Lee was partially supported by the Technology Innovation Program (20011076) of KEIT.
\end{acks}

\bibliographystyle{ACM-Reference-Format}
\bibliography{reference}

\pagebreak

\appendix
\renewcommand*\appendixpagename{Supplementary Material}
\renewcommand*\appendixtocname{Supplementary Material}

\begin{appendices}
\section{Details of Stage-2 training}
Algorithm 1 shows the procedure of training the NCM in stage-2.
Given a short sequence of source context feature $z_{src}^i|_{i=0}^s$, we find the target feature sequence  $z_{tar}^i|_{i=0}^s$ from $\mZ_{tar}$ such that $f(z_{tar}^0)$ is closest to $f(z_{src}^0)$. We argue that comparing only the first features is sufficient because they already encode 1 second of motion and the sequence length $(s=8)$ is short.
After that, we train the networks in an auto-regressive fashion. 
First, $E_v$ creates a posterior distribution while $p$ creates a prior distribution. We sample a latent variable $s^i$ from the prior distribution. Finally, $D_v$ predicts a matched character feature $\tilde{z}^{i}$. We use two loss terms, $\mathcal{L}_{zval}$ to measure the error and $\mathcal{L}_{kl}$ for KL-divergence losses.

\begin{algorithm}[t]
\DontPrintSemicolon
    \tcc{$\mathcal{D}_{tar} = [\mZ_{tar}, f(\mZ_{tar})]$: target character feature dataset (e.g. Zombie)}
    \tcc{$z^i_{src}|_{i=0}^s$: s-frames sequence of source features}    
    \SetKwFunction{FMain}{Train NCM}
    \SetKwProg{Fn}{Function}{:}{}
    \Fn
    {\FMain{$\theta_{E_v}, \theta_p, \theta_{D_v}$}} 
    {
        $l \gets \text{action label of} \,\, z^0_{src}$ \;
        \tcc{Find nearest neighbor with same action label}
        $z_{cha}^0 \gets \text{Nearest}(f(z^0_{src}), f(\mZ_{tar});l)$\;
        \tcc{Set $s$-frames sequence of character features starting from $z_{cha}^0$ as prediction targets}
        $z_{cha}^i|_{i=0}^s \in \mZ_{tar}$ \;
        \tcc{Set initial state for prediction}
        $\tilde{z}^0 \,\, \gets \,\, z_{tar}^0$\;
        \tcc{Predict feature sequence $\tilde{z}^{i}|_{i=1}^s$}
        \For{$i \,\, \gets \,\, 1$ to $s$} {
            \tcc{Predict $\mu_{po}$,$\sigma_{po}$ to create posterior dist}
            $\mu_{po}, \, \sigma_{po} \,\, \gets \,\, E_v(z_{cha}^{i}, \tilde{z}^{i-1}, f(z_{src}^i); \theta_{E_v})$\;
            \tcc{Predict $\mu_{pr}$,$\sigma_{pr}$ to create prior dist}
            $\mu_{pr}, \, \sigma_{pr} \,\, \gets \,\, p(\tilde{z}^{i-1}, f(z_{src}^i); \theta_{p})$\;
            \tcc{Sample latent variable $s^i$}
            $ s^i \,\, \gets \,\, \mathcal{N}(\mu_{pr}, \sigma_{pr})$ \;
            \tcc{Predict matched feature $\tilde{z}^{i}$}
            $\tilde{z}^{i} \,\, \gets \,\, D_v(s^i, \tilde{z}^{i-1}, f(z_{src}^i); \theta_{D_v})$\;

            \tcc{Compute losses}
            $\mathcal{L}_{zval} \,\, \gets \,\, \|z^i_{cha} - \tilde{z}^i \|_1 ,$\;
            $\mathcal{L}_{kl} \,\, \gets \,\, \text{KL divergence}(\mu_{po},\sigma_{po} \| \mu_{pr},\sigma_{pr})$\;
            \tcc{Update network parameters}
            $\theta_{E_v}, \theta_p, \theta_{D_v} \,\, \gets \,\, \text{AdamW}(\theta_{E_v}, \theta_p, \theta_{D_v}, \mathcal{L}_{zval}+\lambda_{kl}\mathcal{L}_{kl}) $\;
        }
    }
    \textbf{End}
\caption{Training for Neural Context Matching $\mathcal{NCM}$}
\label{alg:nueral_context_matching}
\end{algorithm}

\section{Implementation details}
\label{sec:implementation}
The AdamW optimizer was used over 125 epochs for stage-1 and 30,000 iterations for stage-2, with a learning rate scheduler set to decay at a rate of $1e^{-4}$.
Loss weights were set as: $\lambda_{id}=1$, $\lambda_{cyc}=1$, $\lambda_{ctr}=0.1$, $\lambda_{kl}=1$. 
Every transformer-based network comprised 2 layers of 256 channels with 4 heads, and the context mapping function $f(\cdot)$ was modeled with 2 MLP layers followed by Instance Normalization.
With two 12GB 2080ti GPUs, stage-1 training took around 20 hours per character while stage-2 training took a full day per character.

\section{Additional Ablation Study}
\label{sec:additional_ablation}
We performed further ablation studies to highlight the efficacy of the Prior Net and the autoregression scheme of the Neural Context Matcher.

\begin{figure}[t]
  \centering
  \includegraphics[width=0.95\linewidth]{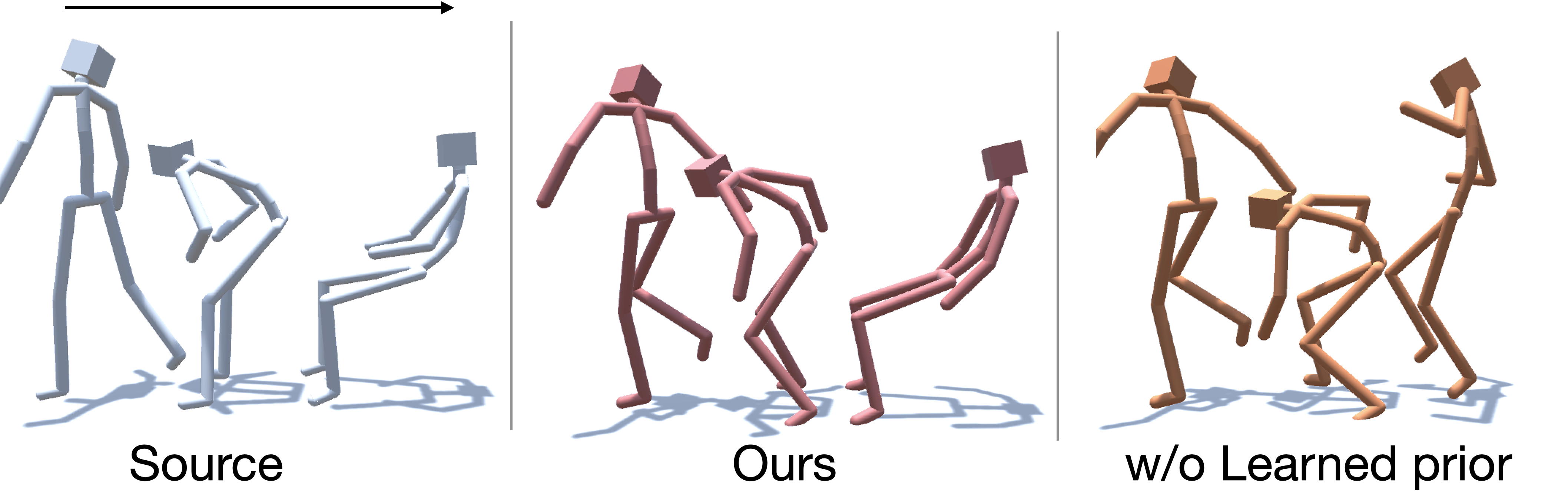}
  \caption{Ablation test on Prior Net. Result obtained without learned prior distribution fails to generate Neutral Clown sitting pose.}
  \label{fig:ablation_prior}
\end{figure}

\paragraph{\textbf{Effect of Prior Net}} 
The Prior Net in the Neural Context Matcher plays an important role in generating the current character feature $z_{cha}^{i}$ from the sampled latent variable $s^i$ by configuring a prior distribution that is appropriate for the previous character feature $z_{cha}^{i-1}$ and the current source feature $f(z_{src}^{i})$. To assess its impact, we trained a Neural Context Matcher without the Prior Net, sampling latent variables from a $\mathcal{N}(0,1)$ Gaussian distribution.
As Figure~\ref{fig:ablation_prior} shows, when the ablated model was used to characterize a sitting motion as Neutral Clown, it struggled to generate a motion that corresponded to the context feature of sitting.

\begin{figure}[t]
  \centering
  \includegraphics[width=0.95\linewidth]{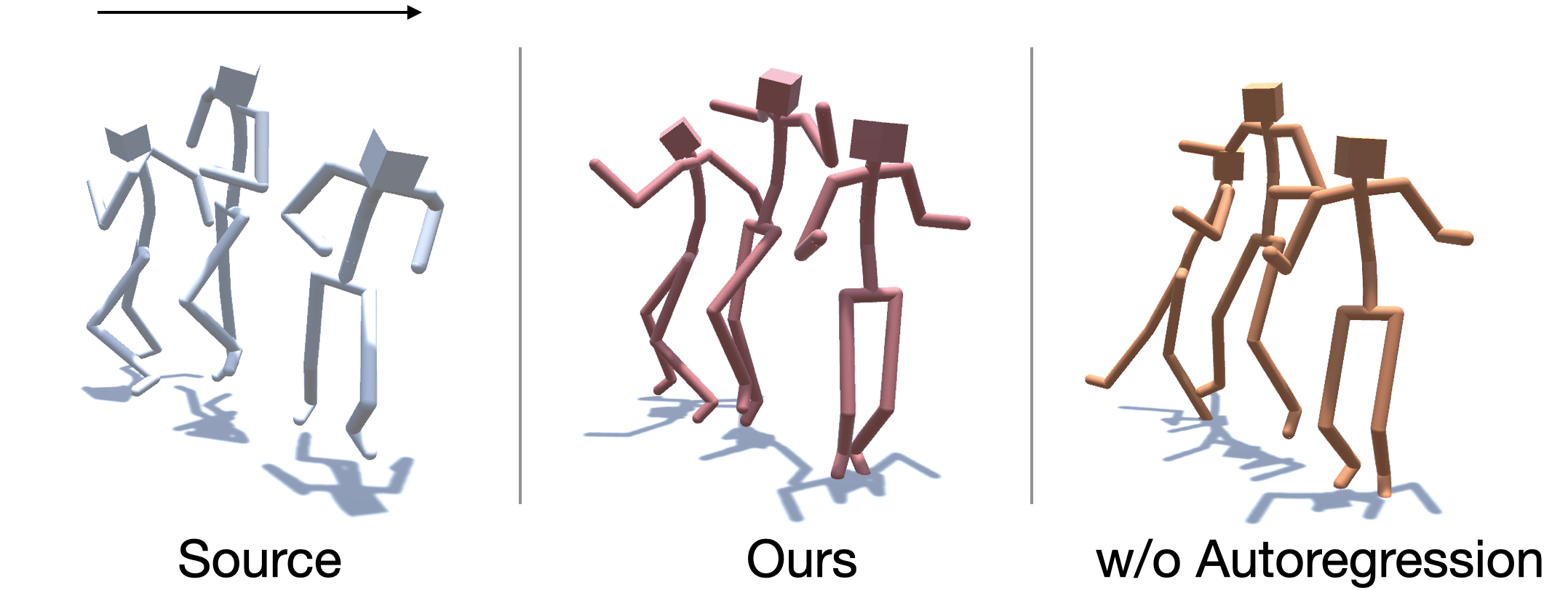}
  \caption{Ablation test on the autoregressive scheme of the NCM. Our characterized output motion achieves higher quality of temporal consistency and transition naturalness compare to what is obtained without autoregressively feeding the previous character feature.}
  \label{fig:ablation_autoregression}
\end{figure}

\paragraph{\textbf{Effect of autoregression}}
To examine the effect of our autoregression framework, we tested the stage-2 training by removing the condition of the previous character feature $\tilde{z}^{i-1}$ from the Prior Net and the NCM Decoder (Al.~\ref{alg:nueral_context_matching}). 
Figure~\ref{fig:ablation_autoregression} shows snapshots of characterizing a turn-and-run motion as Neutral Clown. The ablated model, which does not take into account the previous character feature, not only created evident discontinuities between poses but also failed to generate a smooth transition between turn and run. In contrast, our model was capable of generating continuous and plausible characterized motion.

\section{Comparison with Projector of LMM}
\label{sec:projector}
\begin{figure}[t]
  \centering
  \includegraphics[width=0.95\linewidth]{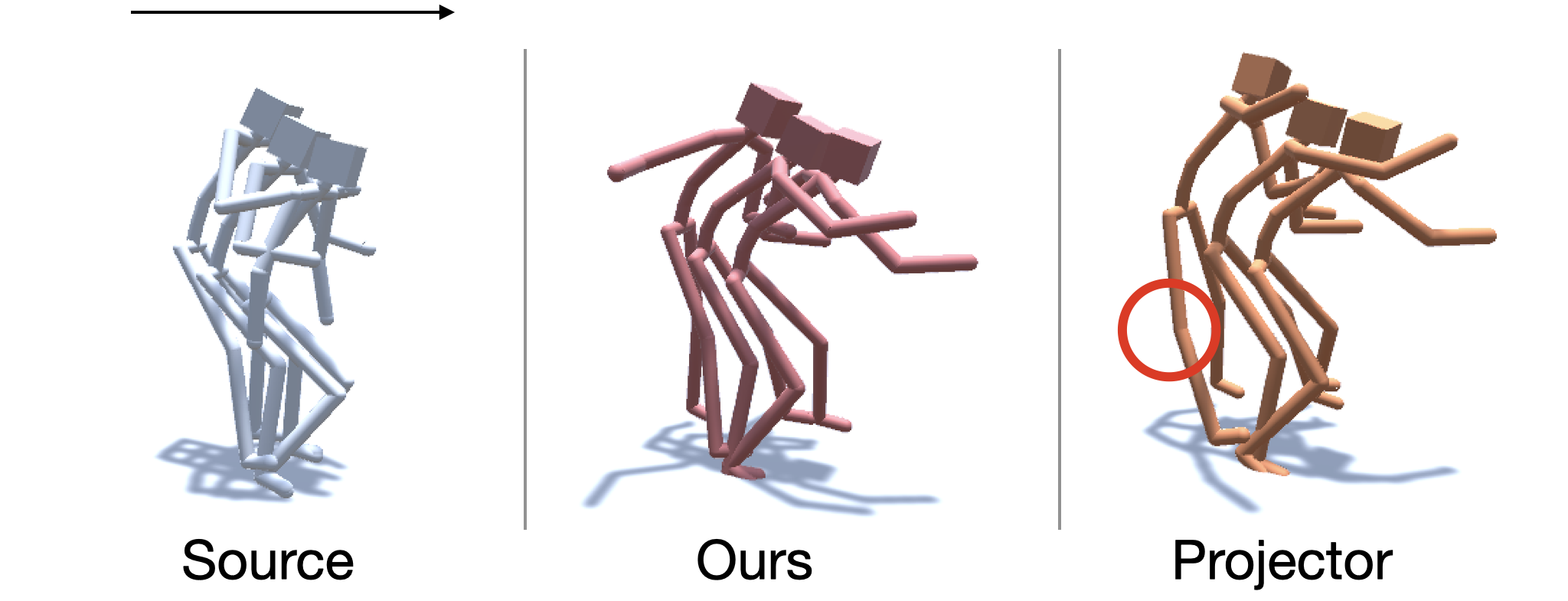}
  \caption{Comparison with the Projector of the Learned Motion Matching. The knee joint of the output pose is bent backwards, creating an implausible pose (red circle).}
  \label{fig:comparison_projector}
\end{figure}

We conducted a comparison with a model that replaces our NCM with the Projector network demonstrated in the Learned Motion Matching technique [Holden et al. 2020]. Unlike our model, the Projector network does not consider temporal continuity or the prior distribution, which may lead to the loss of temporal consistency or the generation of implausible poses.
Figure \ref{fig:comparison_projector} shows an example of characterizing a motion of landing-and-preparing-for-jump as Happy Zombie, in which the resulting motion generated by the model based on the Projector did not reflect the source motion's context well and also had an unnatural knee bending.

\end{appendices}

\end{document}